\documentclass[submission, Proceedings]{SciPost}

\hypersetup{
    colorlinks,
    linkcolor={red!50!black},
    citecolor={blue!50!black},
    urlcolor={blue!80!black}
}

\usepackage[bitstream-charter]{mathdesign}
\DeclareSymbolFont{usualmathcal}{OMS}{cmsy}{m}{n}
\DeclareSymbolFontAlphabet{\mathcal}{usualmathcal}

\usepackage{xspace}
\newcommand{\ft}{\ensuremath{\text{f}^\gamma_\text{T}}\xspace}
\newcommand{\figwidth}{0.43\textwidth}
\newcommand{\whizard}{\textsc{Whizard}\xspace}
\newcommand{\delphes}{\textsc{Delphes}\xspace}

\begin{document}

\begin{center}{\Large \textbf{
      Dark matter searches with mono-photon signature\\
      at future e$^+$e$^-$ colliders
}}\end{center}

\begin{center}
  Jan Kalinowski\textsuperscript{1},
  Wojciech Kotlarski\textsuperscript{2},
  Krzysztof Mekala\textsuperscript{1}, 
  Pawel Sopicki\textsuperscript{1} and\\
  Aleksander Filip \.Zarnecki\textsuperscript{1$\star$}
\end{center}

\begin{center}
  {\bf 1} Faculty of Physics, University of Warsaw \\
  {\bf 2} Institut f\"ur Kern- und Teilchenphysik, TU Dresden \\
* filip.zarnecki@fuw.edu.pl
\end{center}

\begin{center}
\today
\end{center}


\definecolor{palegray}{gray}{0.95}
\begin{center}
\colorbox{palegray}{
  \begin{tabular}{rr}
  \begin{minipage}{0.1\textwidth}
    \includegraphics[width=22mm]{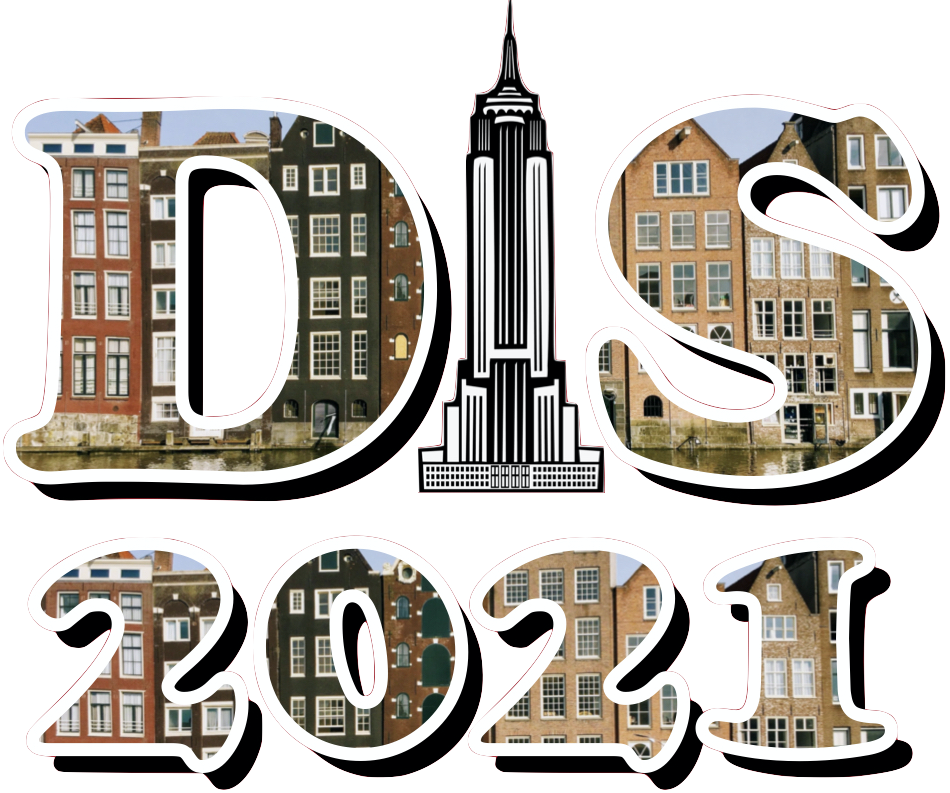}
  \end{minipage}
  &
  \begin{minipage}{0.75\textwidth}
    \begin{center}
    {\it Proceedings for the XXVIII International Workshop\\ on Deep-Inelastic Scattering and
Related Subjects,}\\
    {\it Stony Brook University, New York, USA, 12-16 April 2021} \\
    \doi{10.21468/SciPostPhysProc.?}\\
    \end{center}
  \end{minipage}
\end{tabular}
}
\end{center}

\section*{Abstract}
         {\bf
As any e$^+$e$^-$ scattering process can be accompanied by a hard
photon emission from the initial state radiation, the analysis of the
energy spectrum and angular distributions of those photons can be used
to search for hard processes with an invisible final state. Thus  high
energy e$^+$e$^-$ colliders offer a unique possibility for the most
general search of Dark matter based on the mono-photon signature.   

We consider production of DM particles  via a mediator at the
International Linear Collider (ILC) and Compact Linear Collider (CLIC)
experiments taking into account detector effects  within the \delphes
fast simulation framework.  Limits on the light DM production in a
generic model are set   for a wide range of mediator masses and
widths. For mediator masses up to the centre-of-mass energy of the
collider, results from the mono-photon analysis are more stringent
than the limits expected from  direct resonance searches in Standard Model decay
channels. 
}


\section{Introduction}
\label{sec:intro}

The most general approaches to search for the dark matter (DM) production at
future e$^+$e$^-$ colliders is based on the mono-photon signature,
which is expected when the production of the invisible final state is
accompanied by a hard photon from initial state radiation (ISR).
We proposed the procedure\cite{Kalinowski:2020lhp}
which allows for consistent, reliable simulation of mono-photon events
in \whizard
for both Beyond the Standard Model (BSM) signal
and Standard Model (SM) background processes, based on  merging the matrix
element calculations with the lepton ISR structure function.
For precise kinematic description of photons entering the
detector, emission of up to three photons is included in the matrix
element (ME) calculation. 
Soft and collinear photons are simulated with the \whizard built-in lepton ISR
structure function and a dedicated ISR rejection procedure is applied
to avoid double counting, removing all events with any of the ISR
photons entering the ME photon phase space.
In this contribution, we exploit the proposed procedure
\cite{Kalinowski:2020lhp}  in estimating the
sensitivity of future e$^+$e$^-$ colliders to different DM
scenarios in which DM particles couple to the SM ones via a mediator.
We  propose a novel approach, where the experimental sensitivity is
defined in terms of both the mediator mass and mediator width.
This approach is more model independent than the approaches presented
so far, which assume given mediator coupling values to SM and DM particles
\cite{Habermehl:2020njb,Blaising:2021vhh}. 

\section{Analysis framework}

We consider DM pair-production for ILC running at 500\,GeV,
with total integrated luminosity of 4\,ab$^{-1}$
\cite{Bambade:2019fyw}, and for CLIC running at 3\,TeV, with total
integrated luminosity of 5\,ab$^{-1}$ \cite{Aicheler:2019dhf}. 
Polarisation of both electron and positron beams is considered for the ILC
while only electron beam polarisation is included for CLIC . 

\begin{figure}[tb]
  \centering
  \includegraphics[width=\figwidth]{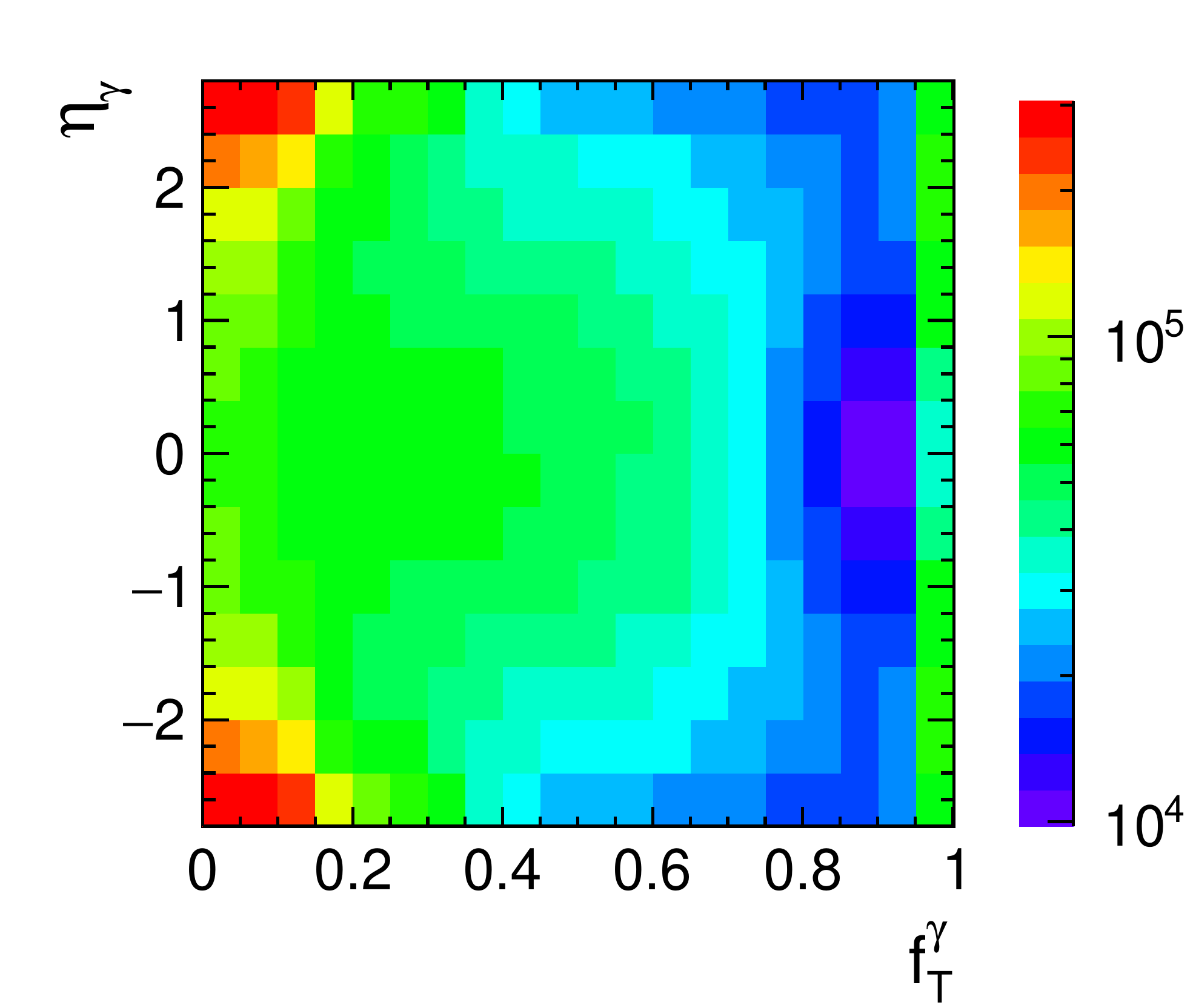}
  \includegraphics[width=\figwidth]{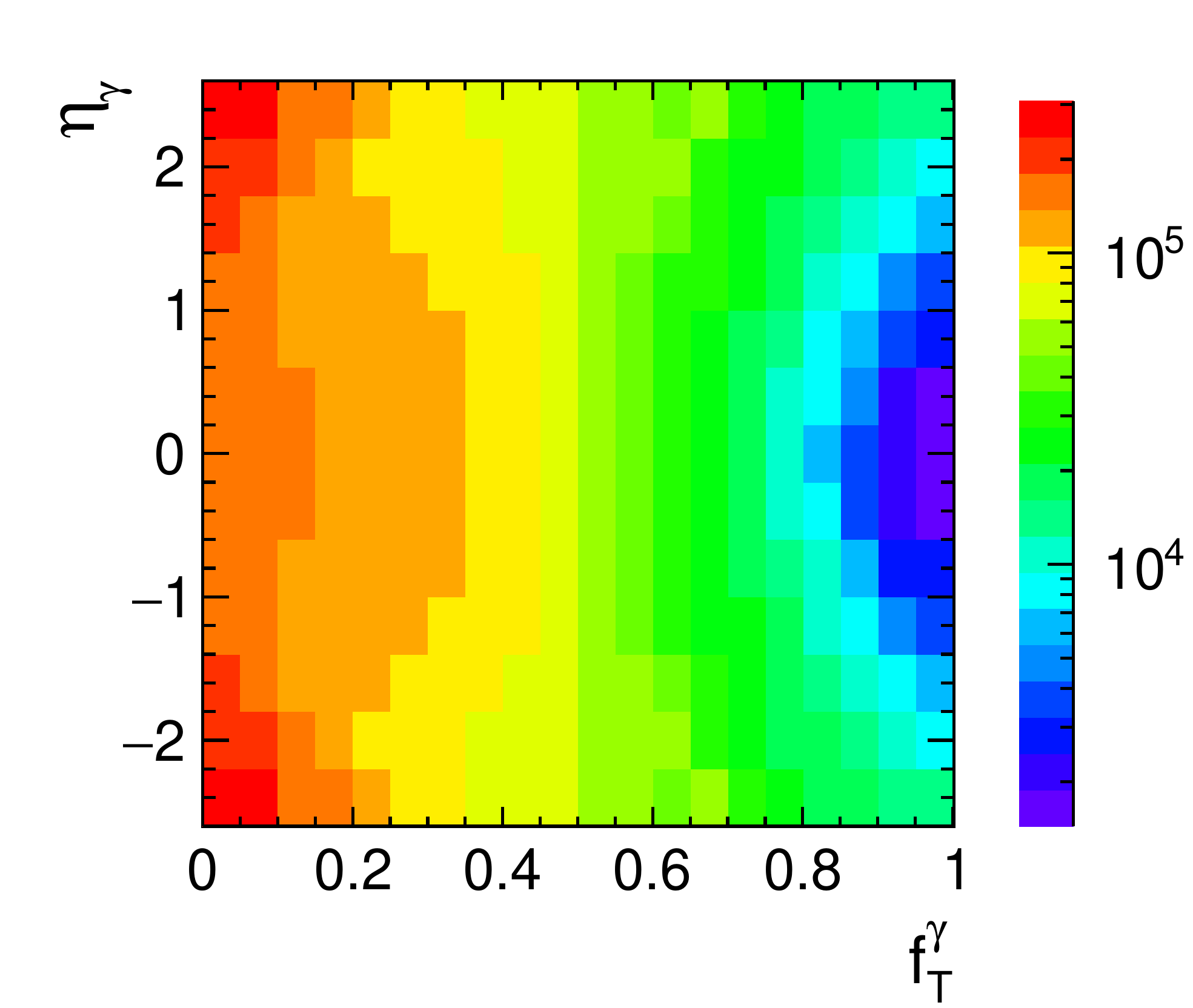}
  \caption{Distribution of the Standard Model background events in the
    $(\ft,\eta_\gamma)$ plane for 500\,GeV ILC running with --80\%/+30\%
    electron/positron beam polarisation (left) and 3\,TeV CLIC running
    with --80\% electron beam polarisation (right).}
  \label{fig:bg2d}
\end{figure}

Signal and background samples generated with \whizard are processed
with the fast simulation framework
\delphes\cite{delph} in which the generic ILC and the CLIC detector models were
implemented.
Both models include detailed description of the forward calorimeter
systems which is crucial for proper modeling of the background
constribution.
Limits on the light DM pair-production cross section are set based on
the expected two-dimensional distributions of the reconstructed
mono-photon events in the rapidity and transverse momentum fraction
defined as
$ \ft = {\log \left( \frac{p_T^{\gamma ~~}}{p_T^{min}} \right)}/
   {\log \left( \frac{p_T^{max}}{p_T^{min}} \right)} \, ,  $
where $p_T^{min}$ is the minimum photon transverse momentum required
in the event selection procedure and $p_T^{max}$ is the maximum
transverse momentum of the photon allowed for the given
scattering angle \cite{Kalinowski:2021tyr}.
Expected distributions of the Standard Model (SM) background events, coming
from the radiative Bhabha scattering and radiative neutrino pair
production, are  shown in Figure~\ref{fig:bg2d}.
\begin{figure}[tb]
  \centering
  \includegraphics[width=\figwidth]{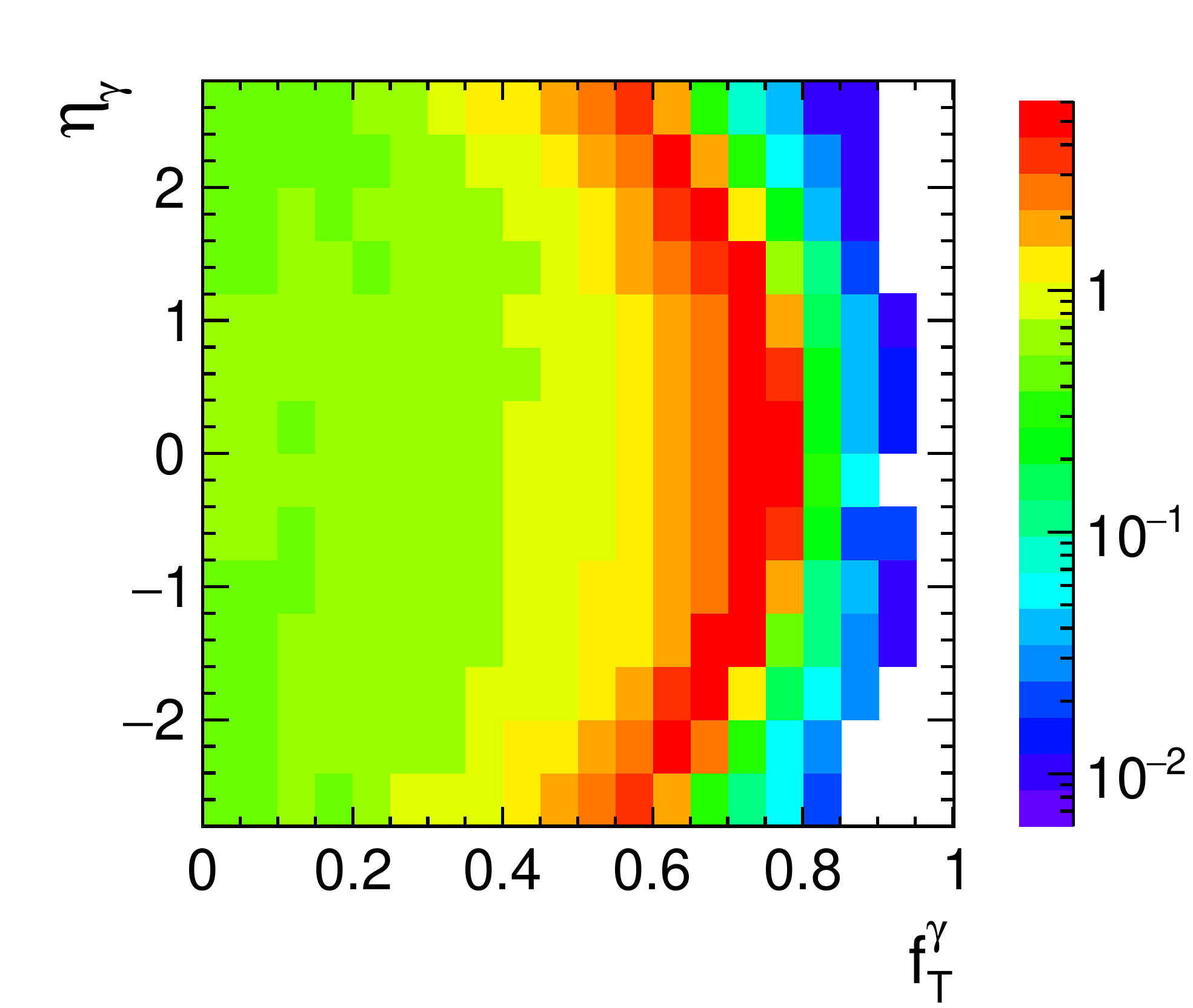}
  \includegraphics[width=\figwidth]{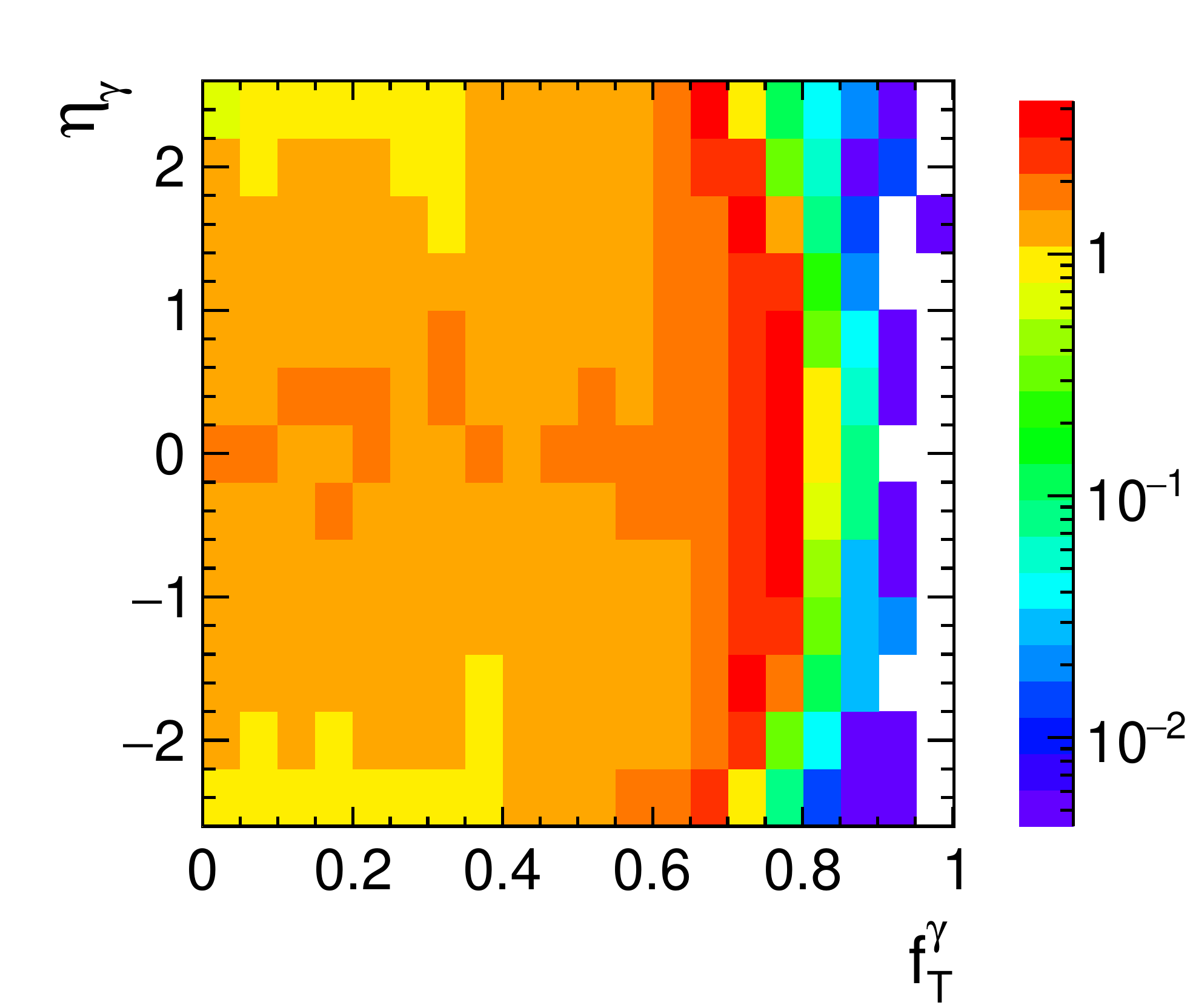}
  \caption{Distribution of the DM pair-production events in the
    $(\ft,\eta_\gamma)$ plane for a fermion DM with M$_\chi$=50\,GeV and
   a vector mediator mass of M$_Y$=400\,GeV at 500\,GeV ILC (left) and
     M$_Y$=2.4\,TeV at 3\,TeV CLIC (right).} 
  \label{fig:sig2d}
\end{figure}
Example distributions of signal events, for a dark matter mass of
M$_\chi$=50\,GeV and mediator mass of M$_Y$=400\,GeV at the ILC and
M$_Y$=2.4\,TeV at CLIC are shown in  Figure~\ref{fig:sig2d}.

\section{Results}

\begin{figure}[tb]
  \centering
  \includegraphics[width=\figwidth]{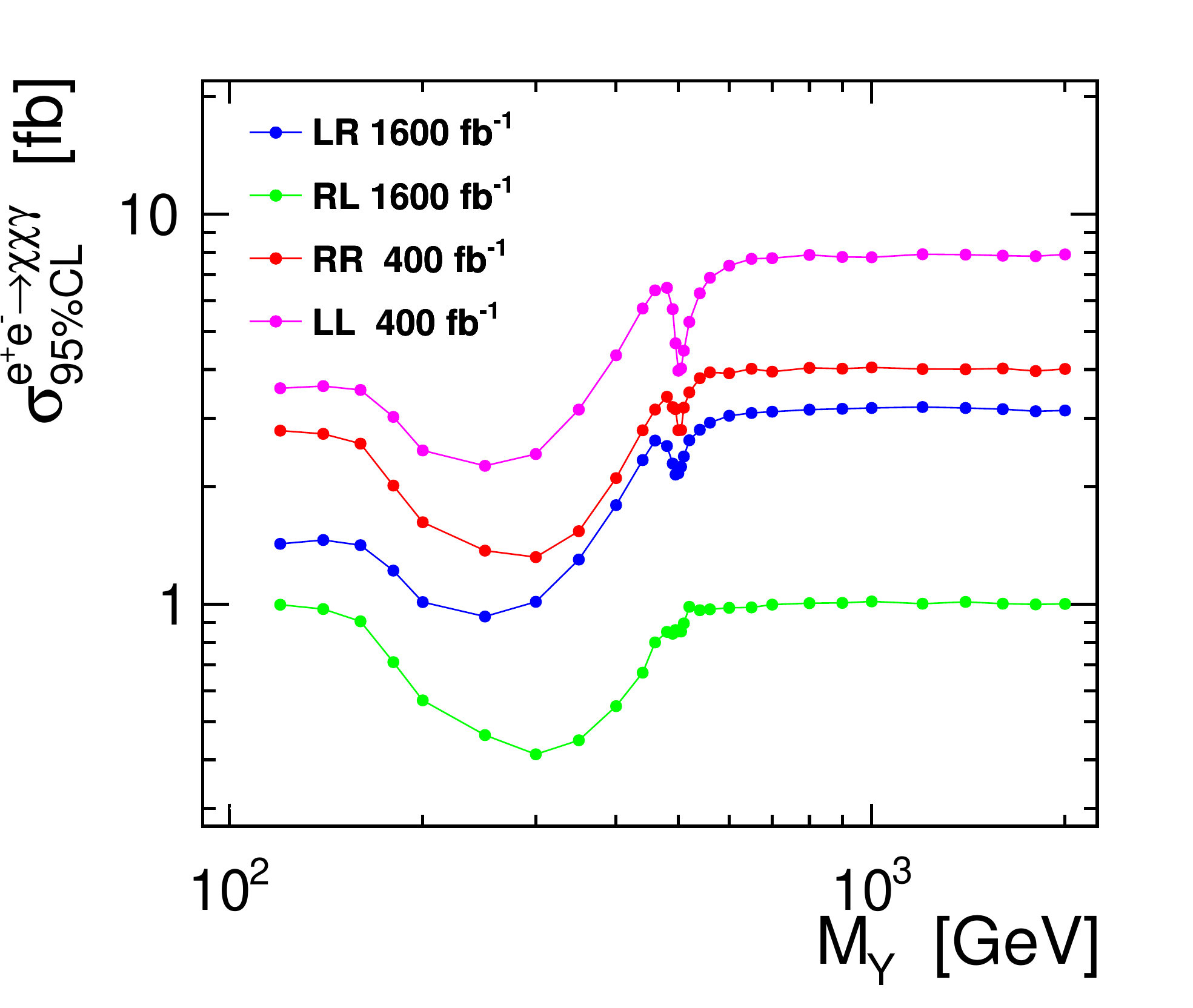}
  \includegraphics[width=\figwidth]{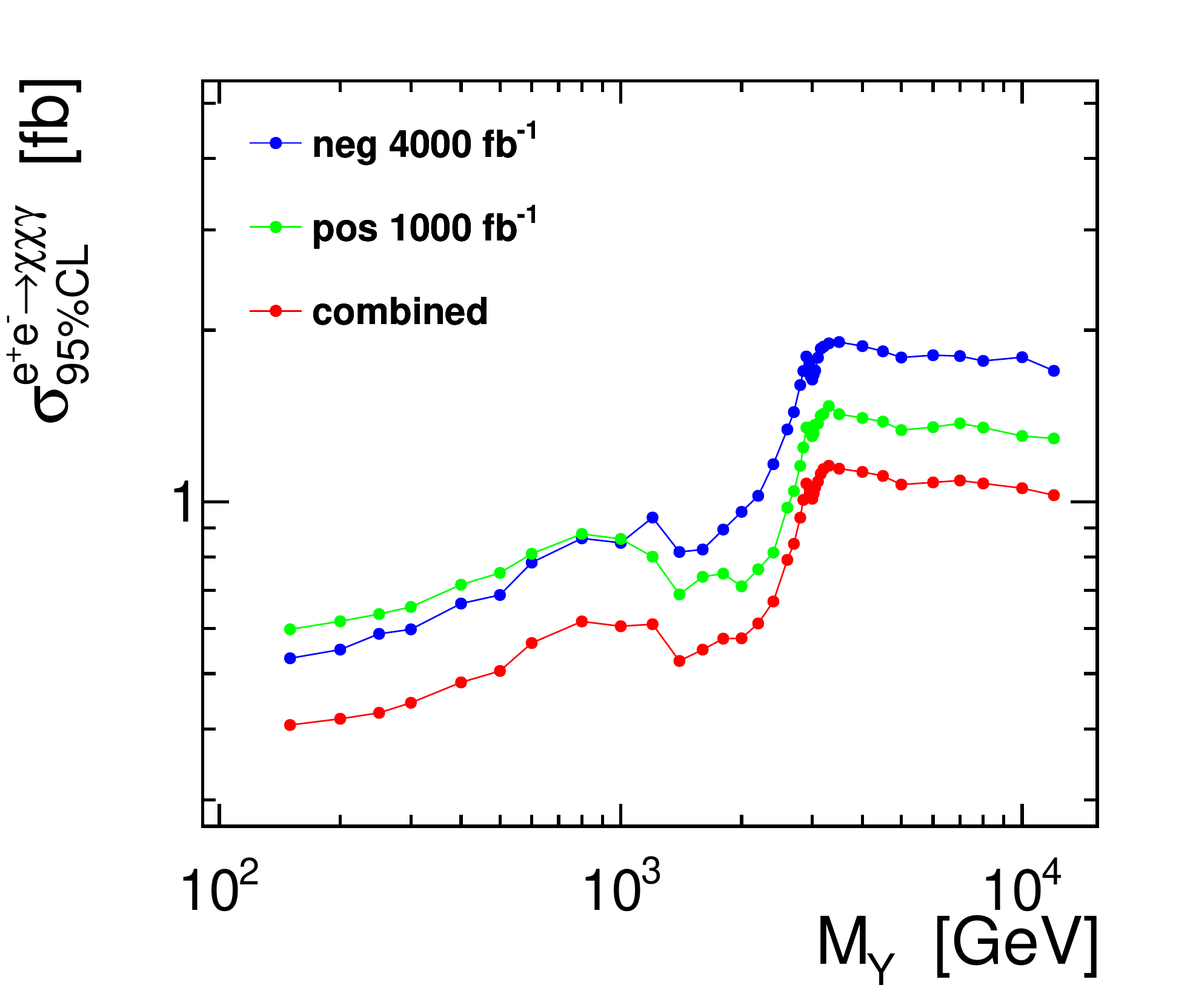}
  \caption{Limits on the radiative DM pair-production cross
    section for processes with the vector mediator exchange, for different
    beam polarisations, at 500\,GeV ILC (left) and 3\,TeV CLIC (right).}
  \label{fig:pol}
\end{figure}

\begin{figure}[tb]
  \centering
  \includegraphics[width=\figwidth]{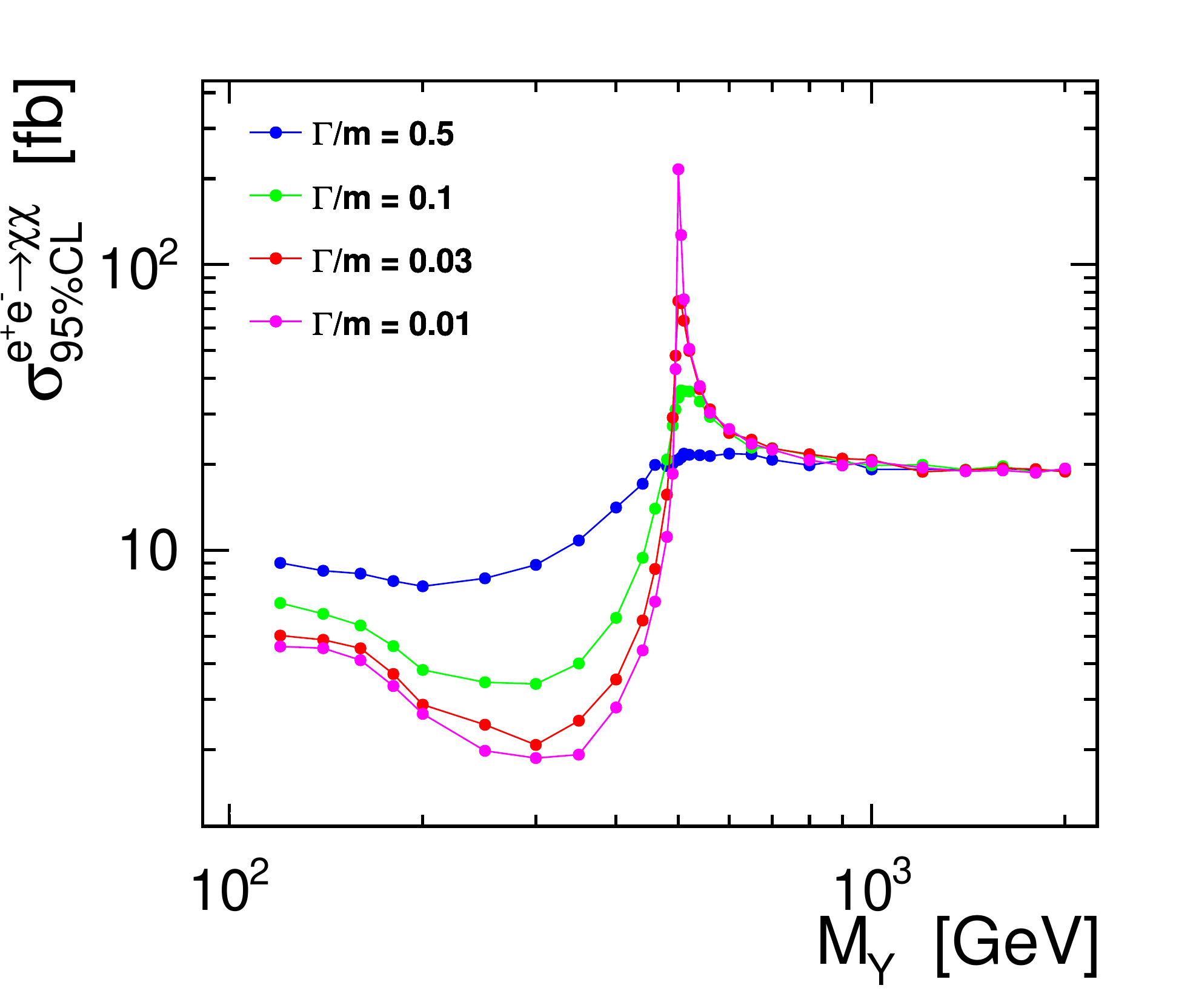}
  \includegraphics[width=\figwidth]{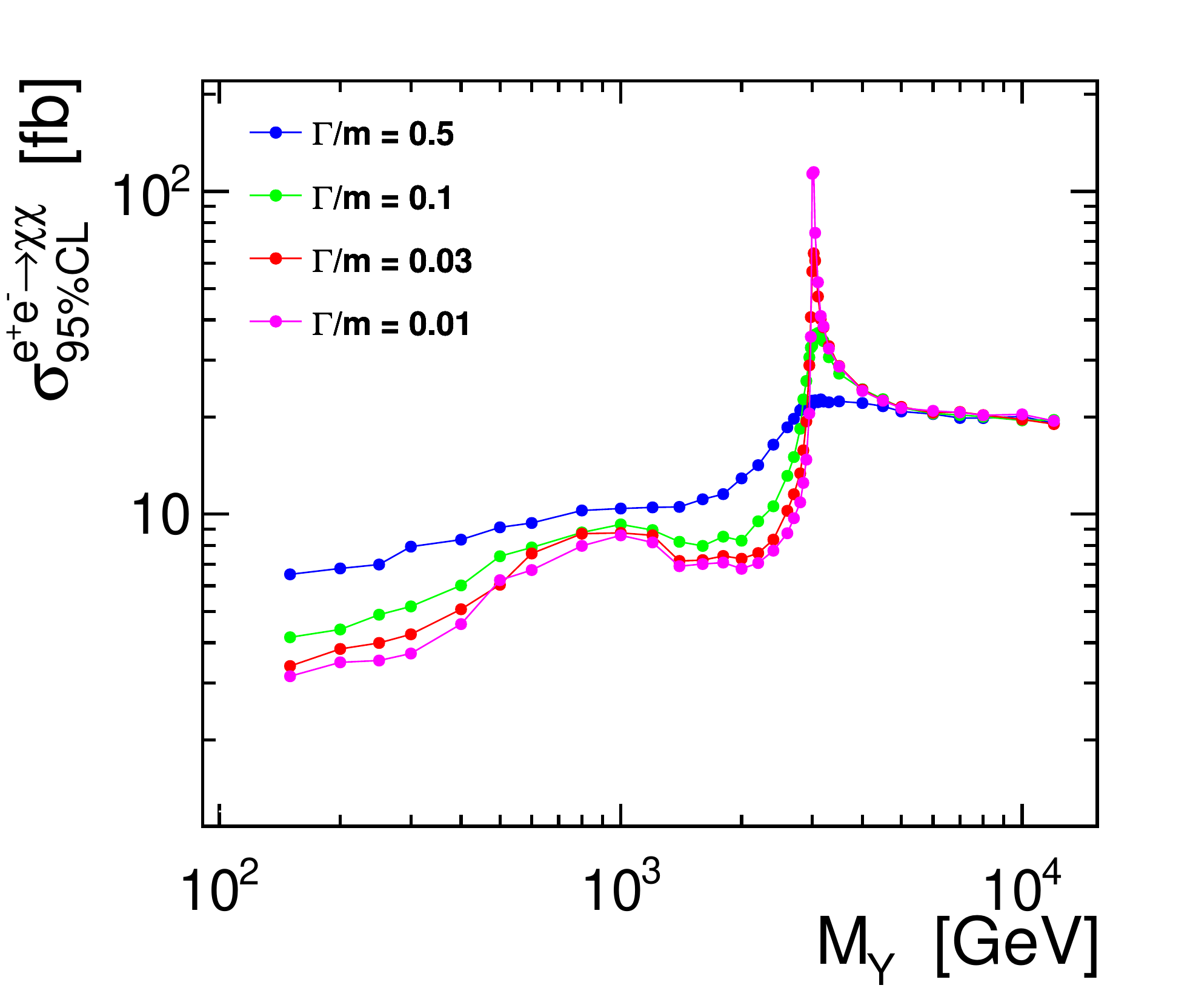}
  \caption{Limits on the total DM pair-production cross
    section for processes with vector mediator exchange, for different
    mediator widths, at 500\,GeV ILC (left) and 3\,TeV CLIC (right).}
  \label{fig:wid}
\end{figure}

Signal and background distributions in the
$(\ft,\eta_\gamma)$ plane are used to extract the 
expected 95\%\,C.L. cross section limits for radiative DM
pair-production. 
Limits for the DM production with exchange of a narrow ($\Gamma$/M=0.03)
vector mediator, expected for the ILC and CLIC running with different beam
polarisations, are shown in Figure~\ref{fig:pol} as a function of the
mediator mass.
While the expected limits strongly depend on the assumed beam
polarisation, both colliders will be sensitive to radiative production
cross sections of the order of 1\,fb for a wide range of mediator masses. 
These results can be corrected for the probability of hard photon
radiation in the detector acceptance region, which significantly
depends on the assumed mediator mass.
It is of the order of 10--15\% for light mediator scenarios and about
5\% for large mediator masses.
However, for mediator masses close to the nominal collision energy,
hard photon radiation is significantly suppressed, in particular for
narrow mediator scenarios.
Resulting limits on the total DM pair-production cross section are
presented in Figure~\ref{fig:wid} as a function of mediator mass, for
different fractional mediator widths.
Total cross section limits are of the order of 10\,fb for the light
mediator, M$_Y < \sqrt{s}$, and about 20\,fb for a heavy one,  M$_Y >
\sqrt{s}$. 
Only for the resonant mediator production, M$_Y \sim \sqrt{s}$, when
the hard photon emission is suppressed, the cross section limits are
much weaker. 

\begin{figure}[tb]
  \centering
  \includegraphics[width=\figwidth]{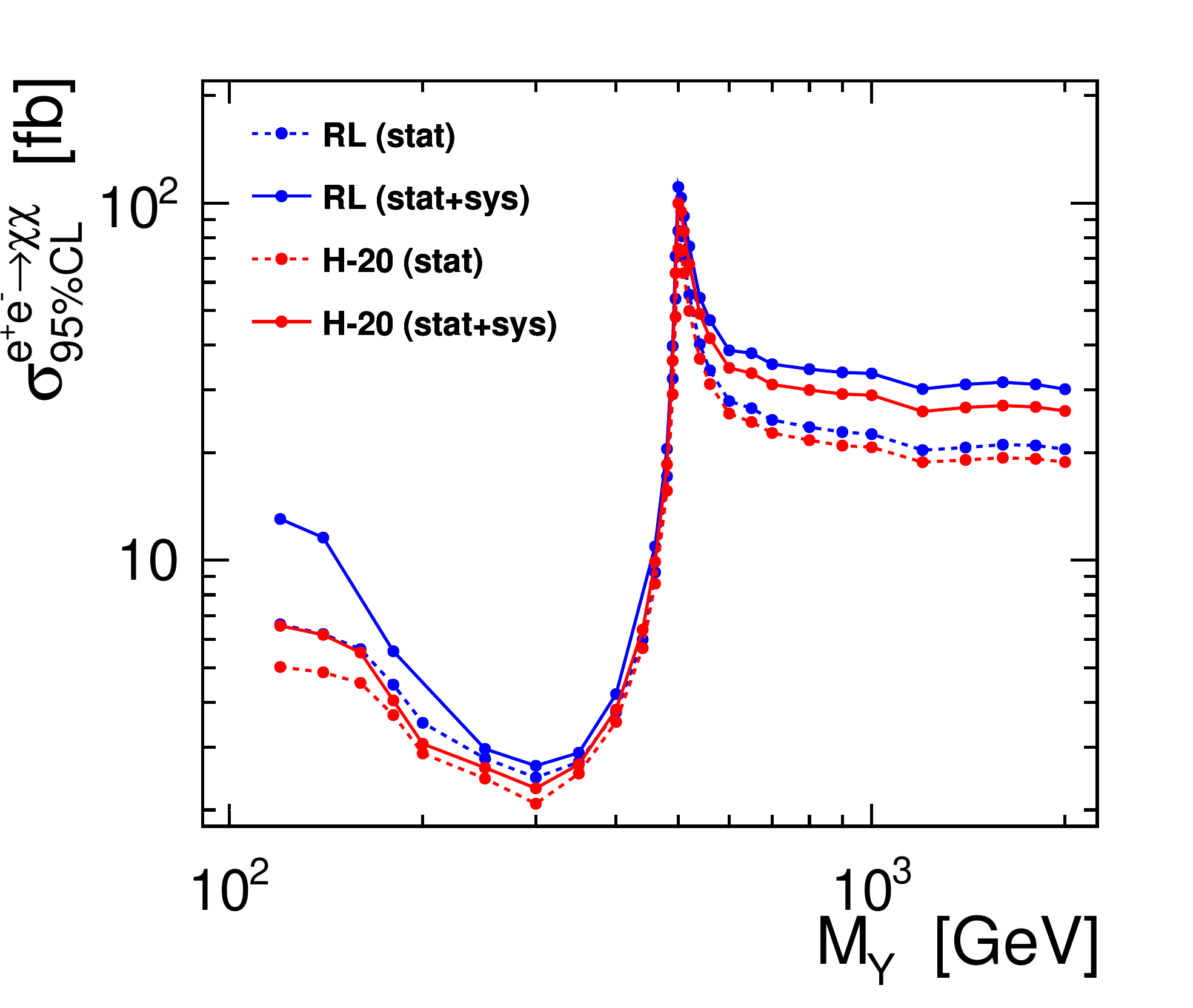}
  \includegraphics[width=\figwidth]{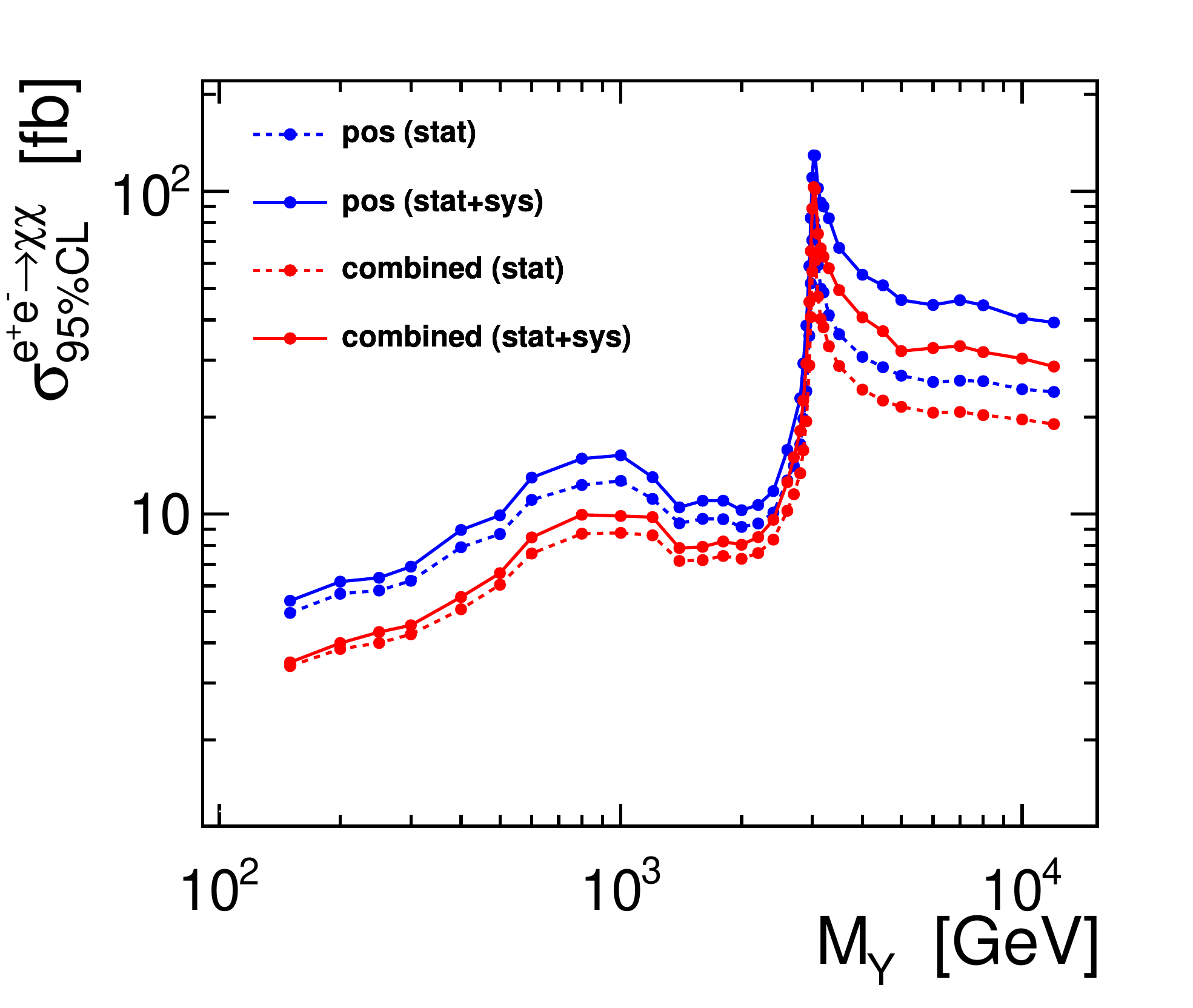}
  \caption{Limits on the total DM pair-production cross
    section for vector mediator width, $\Gamma$/M=0.03, at 500\,GeV
    ILC (left) and 3\,TeV CLIC (right) without (stat) and
    with (stat+sys) systematic uncertainties taken into account.}
  \label{fig:sys}
\end{figure}

The impact of systematic uncertainties was considered following the
approach presented in \cite{Habermehl:2020njb} and the results are
shown in Figure~\ref{fig:sys} for a scenario with a narrow vector mediator
exchange.
While systematic uncertainties deteriorate the expected cross section
limits by about 50\% for heavy mediator scenarios, their impact is 
significantly reduced for light mediator exchange, when a
resonance-like structure is expected in the mono-photon spectra (see
Figure~\ref{fig:sig2d}).

\begin{figure}[tb]
  \centering
  \includegraphics[width=\figwidth]{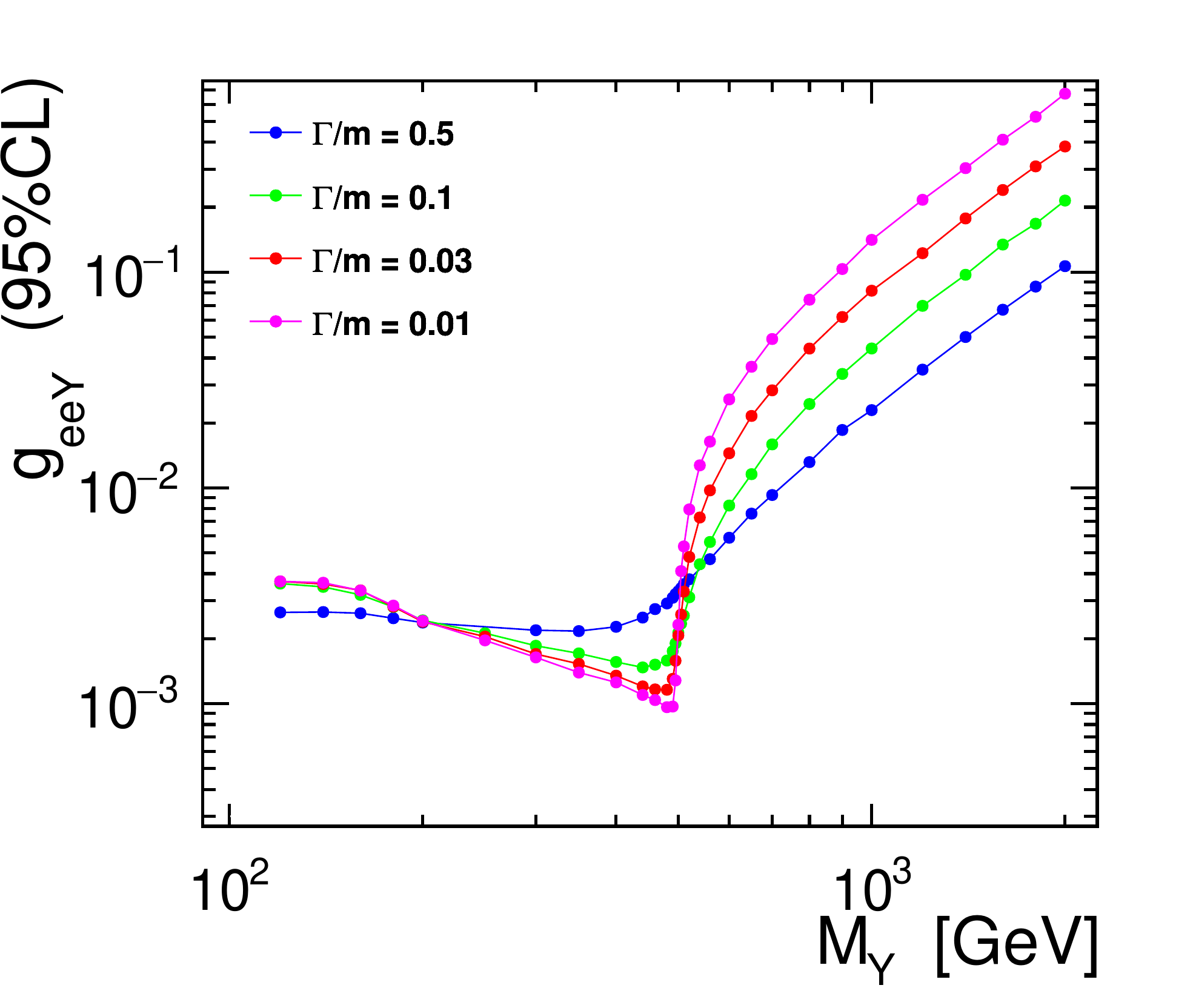}
  \includegraphics[width=\figwidth]{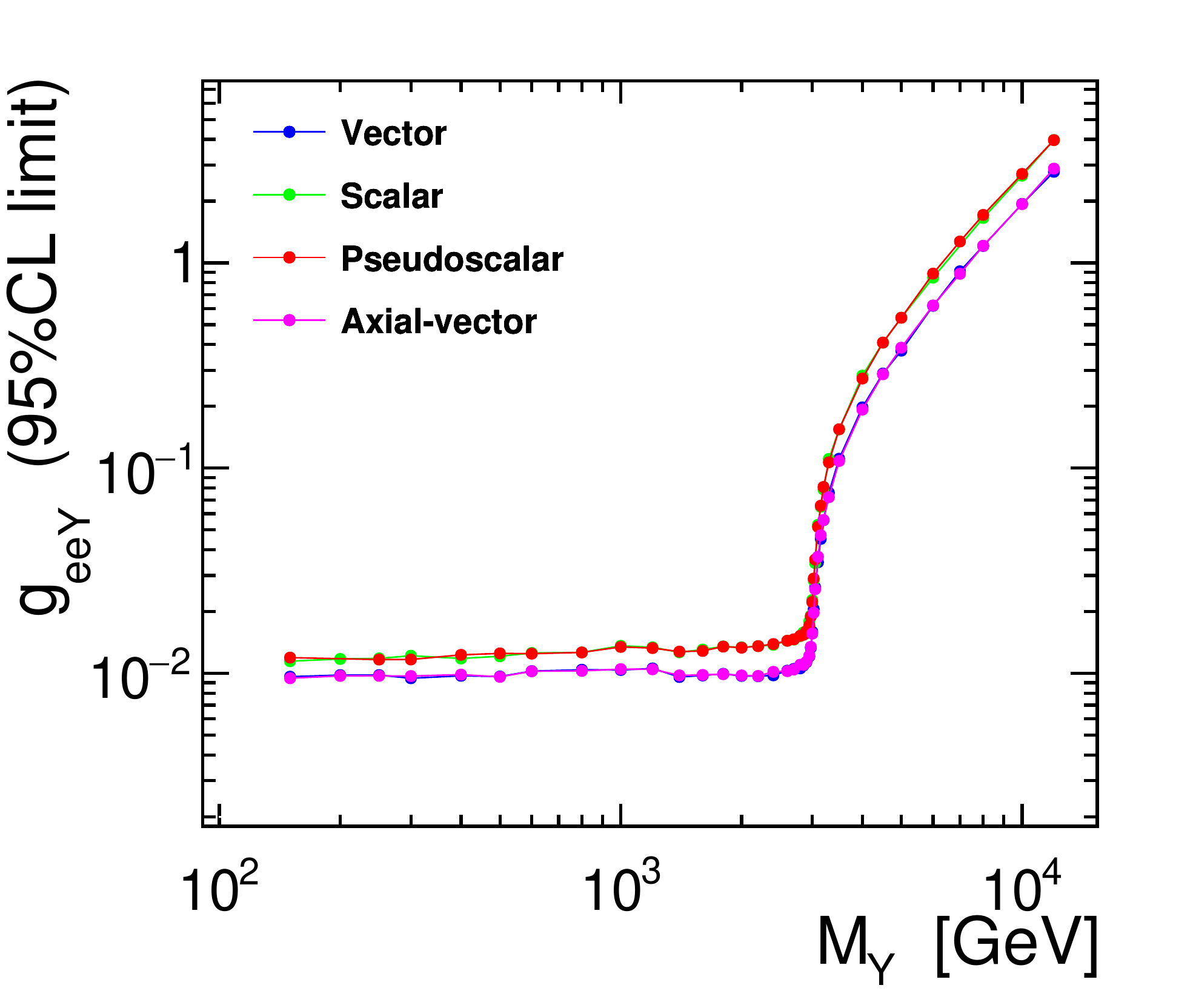}
  \caption{Limits on the mediator coupling to electrons for different
    relative widths of vector mediator at 500\,GeV ILC (left) and for
    mediator width of $\Gamma$/M=0.03 and different mediator types at
    3\,TeV CLIC (right).} 
  \label{fig:coupling}
\end{figure}
Cross section limits corresponding to the combined analysis of the ILC
or CLIC data taken with different beam polarisations, after taking
systematic uncertainties into account, were used to extract the
expected limits on the mediator coupling to electrons, g$_{eeY}$.
Results are presented in Figure~\ref{fig:coupling} for different
vector mediator widths at 500\,GeV ILC and for different mediator
couplings at 3\,TeV CLIC.
The sensitivity of e$^+$e$^-$ colliders to  g$_{eeY}$ is almost uniform up
to the kinematic limit, M$_Y \le \sqrt{s}$.
Coupling limits are about 0.003 at the ILC and about 0.01 at CLIC, and
weakly depend on the assumed coupling structure. 

\section{Conclusion}

Mono-photon production at e$^+$e$^-$ colliders is sensitive
to a wide range of DM pair-production scenarios.
A new framework for mono-photon analyses
\cite{Kalinowski:2021tyr} focuses on scenarios with light mediator
exchange and very small mediator couplings to SM particles.
Limits of the order of 1\,fb and 10\,fb are expected from the
mono-photon analysis for the radiative DM pair-production cross
section and for the total DM pair-production crossection,
respectively, at both the ILC and CLIC, except for the resonance region,
 M$_Y \approx \sqrt{s}$.
 Limits on the mediator coupling to electrons in the range
 $10^{-3}-10^{-2}$ are obtained up to the kinematic limit.
 These limits are more stringent than those expected from direct
resonance search in SM decay channels.

\section*{Acknowledgements}


We thank members of the CLIC detector and physics (CLICdp) collaboration
and the International Large Detector (ILD) concept group for the ILC
for fruitful discussions, valuable comments and suggestions. 
%
This contribution was supported by the National Science Centre, Poland,
the OPUS project under contract UMO-2017/25/B/ST2/00496 (2018-2021) and
the HARMONIA project under contract UMO-2015/18/M/ST2/00518 (2016-2021),
and by the German Research Foundation (DFG) under grant number STO 876/4-1
and STO 876/2-2.



\bibliography{dis2021_dm.bib}

\nolinenumbers

\end{document}